\documentclass[aps,prb,twocolumn,superscriptaddress,showpacs,floatfix]{revtex4}

\usepackage{graphicx}

\graphicspath{{.}{./eps/}}

\usepackage[T1]{fontenc}
\usepackage{amsmath}
\usepackage{amssymb}
\usepackage{url}
\usepackage{color}

\newcommand{\E}{\ensuremath{{\mathrm{e}}}}
\newcommand{\imag}{\ensuremath{{\mathrm{i}}}}
\newcommand{\greenf}{\ensuremath{\mathcal{G}}}

\DeclareMathOperator{\trace}{Tr}

\begin{document}

%\title{Quantum transport through contacted graphene based structures: Disorder effects}
%\title{Spectral, optical, and transport properties  of disordered graphene nanoribbons}
%\title{Single-particle and optical spectrum of disordered graphene nanoribbons}
\title{Effects of disorder and contacts on transport through  graphene nanoribbons}
\author{A. Pieper}
\affiliation{Institut f\"ur Physik, Ernst-Moritz-Arndt-Universit\"at
  Greifswald, 17487 Greifswald, Germany}
  \author{G. Schubert}
\affiliation{Philips Healthcare, \"Ayritie 4, 01510 Vantaa, Finland}
  \author{G. Wellein}
\affiliation{Regionales Rechenzentrum Erlangen, 
Universit\"at Erlangen-N\"urnberg, 91058 Erlangen, Germany }
%\author{A. R. Bishop}
%{\affiliation{\mbox{Theory, Simulation and Computation Directorate, 
%Los Alamos National Laboratory, Los Alamos, New Mexico 87545}}
\author{H. Fehske}
\affiliation{Institut f\"ur Physik, Ernst-Moritz-Arndt Universit\"at
 Greifswald, 17487 Greifswald, Germany}
\date{\today}
\begin{abstract}
We study the transport of charge carriers through finite graphene structures. The use of numerical exact kernel polynomial and Green function  techniques allows us to treat actual sized samples beyond the Dirac-cone approximation. Particularly we investigate disordered 
nanoribbons, normal-conductor/graphene  interfaces and normal-conductor/graphene/normal-conductor 
junctions with a focus on the behavior of the local density of states, single-particle spectral function, optical conductivity 
and conductance. We demonstrate that the contacts and bulk disorder will have a major impact on
the electronic properties of  graphene-based devices.
 \end{abstract}
\maketitle

%multi-cite e.g. : longitudinal ribbon size~\cite{AGW07,LBNR08,NRC08}.
%

%============================================================================================================

\section{Introduction}

During the last decade graphene and graphene based nanostructures have attracted a 
great amount of attention in regard to both fundamental research and application device engineering. 
Most of the unique  electronic properties of graphene originate 
from the strictly two-dimensional arrangement of carbon atoms on a 
honeycomb lattice  and the related gapless conical low-energy spectrum 
around the corners ($K$ and $K'$ points) of the hexagonal Brillouin zone. 
There bulk graphene supplies charge carriers that have sublattice and valley pseudospins, 
feature the pseudorelativistic dynamics of (massless) Dirac fermions and consequently 
possess a fixed chirality (helicity).  These characteristics lead to many unusual and sometimes
counterintuitive charge transport phenomena such as a finite ``universal'' dc conductivity at the neutrality point, Klein tunneling,  
or an anomalous quantum Hall effect; for a recent review see Ref.~\onlinecite{CGPNG09}.  
Concerning the optical properties, within the Dirac cone approximation, only transitions across the Dirac point that are vertical
in momentum space are allowed, leading to a frequency-independent absorption of undoped graphene.\cite{YRDK11} 
For doped graphene the optical response is greatly reduced for frequencies smaller than twice the absolute 
value of the Fermi energy due to Pauli's exclusion principle, while for larger frequencies it is roughly 
given by an universal ac conductivity.\cite{SPFA13} 
  
Recent breakthroughs in graphene  fabrication and patterning facilitated the realization of graphene based electronics, plasmonics and optics.   
In particular graphene nanoribbons (GNRs) with varying widths down to a few nanometer and graphene quantum dots have been prepared 
and operated with field-effect transistor, filter, polarizer or electronic lens functionalities.  
The striking electronic properties of these GNR based nanostructures are strongly dependent on their geometry and edge shape.\cite{GALMW07,STC08,ZG09}
GNRs with zigzag or armchair shaped edges develop specific band structures.\cite{Ez06,Ez07}   Thereby, for a realistic modeling of 
the GNR's quasiparticle energies and band gaps,  edge passivation, edge closure and edge bond relaxation have to be taken into account.\cite{SCL06,ZG09,KOQF11} 
In narrow  armchair GNRs with hydrogen termination aromatic (Clar) sextets largely affect the band gap and consequently the transport properties.\cite{WSSLM10}
For hydrogen-terminated zigzag GNRs the spin polarization of edge states comes into play.\cite{STC08,GALMW07,WSSLM10}  
Moreover, as a matter of course, the enhanced screened Coulomb interaction gives rise to significant self-energy corrections
for both zigzag and armchair GNRs.\cite{YPSCL07} 
Lastly the leads (contacts) connecting the active graphene element to the electronic reservoirs play always an important role, 
just as the interfaces in graphene junctions and the substrate.

  Regrettably transport through graphene and GNRs  based devices will be strongly affected 
  by disorder,\cite{CGPNG09,ML10} i.e., scattering potentials caused by  intrinsic impurities, bulk defects induced by the substrate, ripples, 
  edge roughness, adsorbent atoms at unsaturated dangling bonds at the boundary of the sample, 
  and adatoms  on graphene's open surface.    
  Disorder is known to be exceedingly efficient in suppressing the charge carrier's mobility in low-dimensional systems, even to the
  point of Anderson localization.  However graphene shows distinctive features in this respect too. First, only short-range impurities may cause
intervalley scattering leading to Anderson localization.\cite{An58} Second, due to the chirality of the charge carriers quantum interference may trigger even 
weak antilocalization.\cite{TKSG09} Third, charge carrier density fluctuations may break up the sample
into electron-hole puddles; mesoscopic transport is then determined  by activated hopping 
or leakage between the puddles.  Recent observations of Coulomb diamondlike features in device conductance suggest that charge transport in GNRs
occurs through quantum dots forming along the ribbon due to a disorder potential induced by charged impurities.\cite{GTG10}

Experimentally important information about the transport and optical properties of (disordered) GNRs comes from
scanning tunneling  microscopy, angle resolved photoemission spectroscopy, (infrared) optical conductivity and conductance
measurements, scanning probe spectroscopy, current flow  and life time measurements.  

Theoretically these quantities are best 
obtained by unbiased numerical approaches which enable---if many-body interaction effects can be neglected---the treatment 
of actual sized contacted GNRs with and without  disorder beyond the simple continuum  Dirac fermion description.    

In this paper we use highly efficient Chebyshev expansion,\cite{WF08} kernel polynomial\cite{WWAF06} and Green function\cite{Dat95} techniques\cite{FUSH13} to  analyze 
the electronic properties of GNRs with zigzag and armchair edges (Sec.~II), as well as   
disordered normal-conductor(graphene)/GNR junctions (Sec.~III). 
To this end, we calculate the local density of states, the single-particle spectral function, the optical conductivity  
and the conductance for different geometries. 
Special attention is paid to disorder effects.  Studying the influence of disorder on the transport behavior of (contacted) GNRs 
a tight-binding approach  is generally accepted to be a first reasonable starting point.\cite{CGPNG09,XX07,RSOF08,MB09,BC10b,YRDK11,Ch11,SSF11,GSHR13p} 
Then the Hamiltonian for this problem can be written as 
  \begin{equation}\label{H_bm}
     {H} =  \sum_{i=1}V_i^{} {c}_i^{\dag} {c}_i^{} 
           -t \sum_{\langle ij \rangle}({c}_i^{\dag} {c}_j^{} + \text{H.c.})\,,
  \end{equation}
where $c_i^{(\dag)}$ is a fermionic annihilation (creation) operator acting on lattice site $i$ of a honeycomb lattice with $N$ sites, 
$\langle ij\rangle$ denotes pairs of nearest neighbors and the site-dependent on-site potentials $V_i$ can take values  appropriate for the system under consideration.

\section{Disordered Graphene Nanoribbons}
\subsection{Local Density of States}
\begin{figure*}[t]
  \centering
  \includegraphics[width=0.29\linewidth,clip]{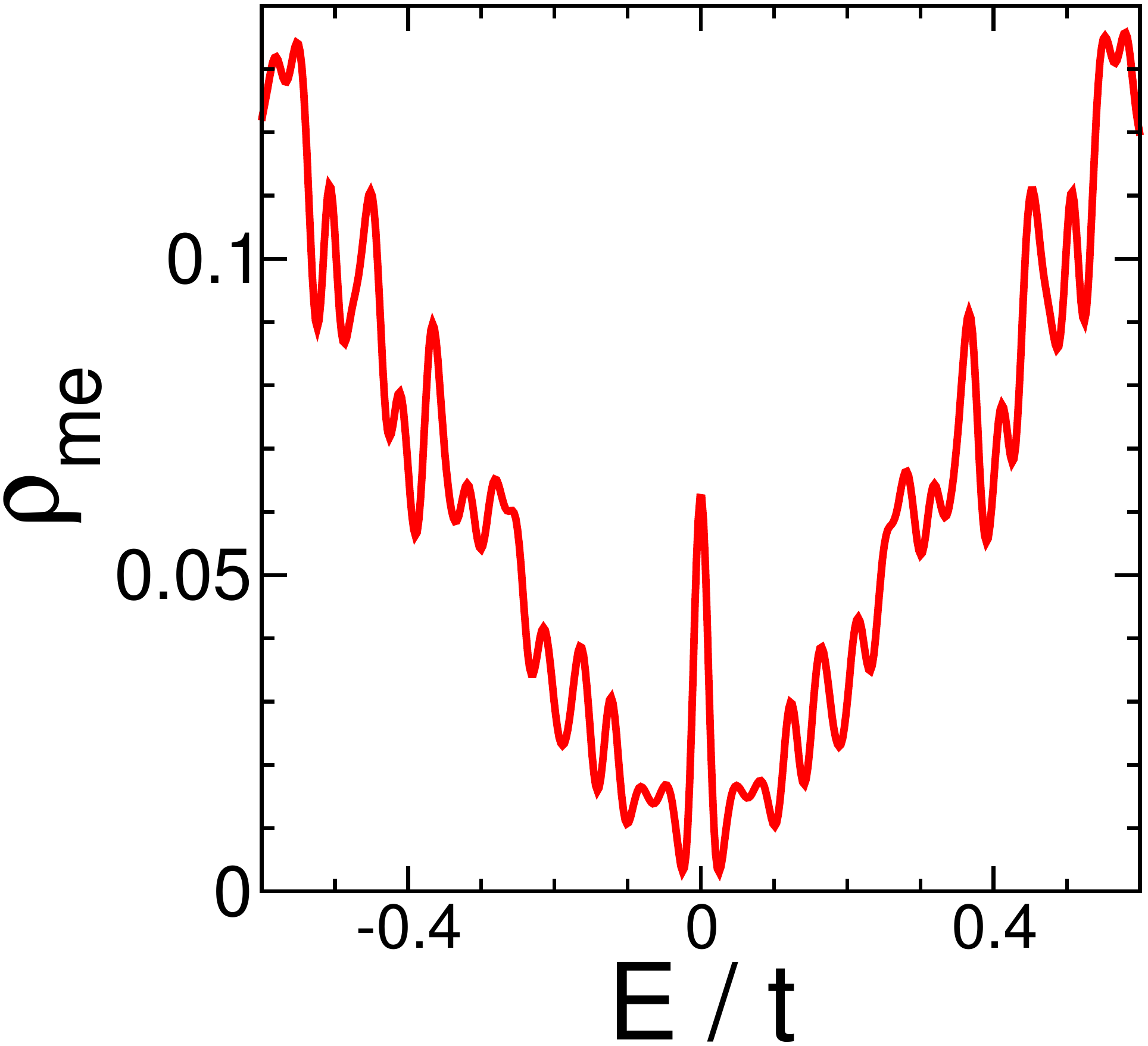}\hspace*{0.6cm}
\includegraphics[width=0.59\linewidth,clip]{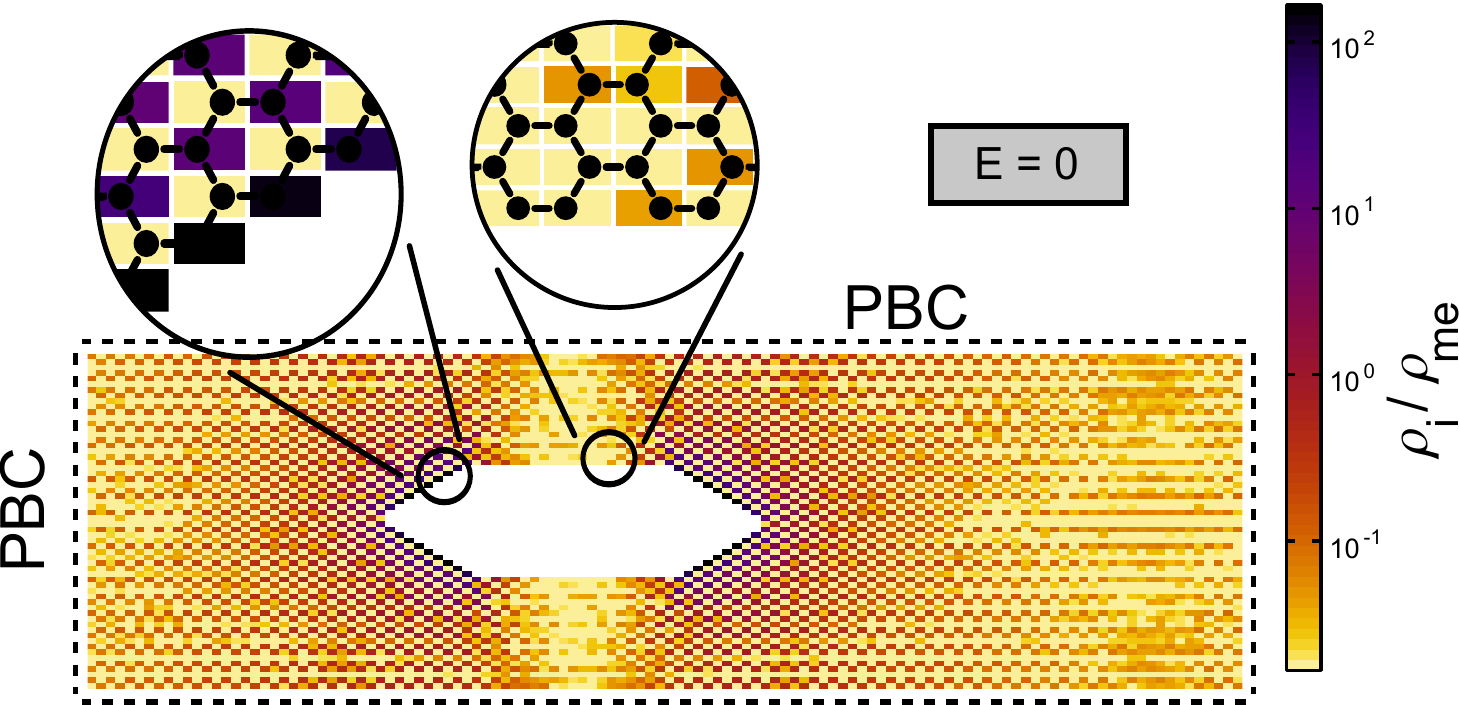}
 \caption{(Color online) Mean DOS  as a function of energy (left panel) and LDOS at $E=0$ (right panel)  for a GNR with 
 $60\times120$ sites and PBC (torus topology) containing a void of unaccessible sites ($V_i=\infty$). Armchair edges are realized at the upper and lower boundary, zigzag edges elsewhere.  Within the KPM calculation of the LDOS $M=2048$ Chebyshev moments were used.
 Note that due to the finite width of the Jackson kernel the KPM assembles contributions
from a couple of eigenstates in the energetic vicinity of the target energy,\cite{WWAF06,SSBFV10}  which, by the way, reflects the situation of a real scanning tunneling microscopy measurement.}
 \label{fig:los_void}
\end{figure*} 
\begin{figure*}[t]
  \centering
  \includegraphics[width=0.3\linewidth,clip]{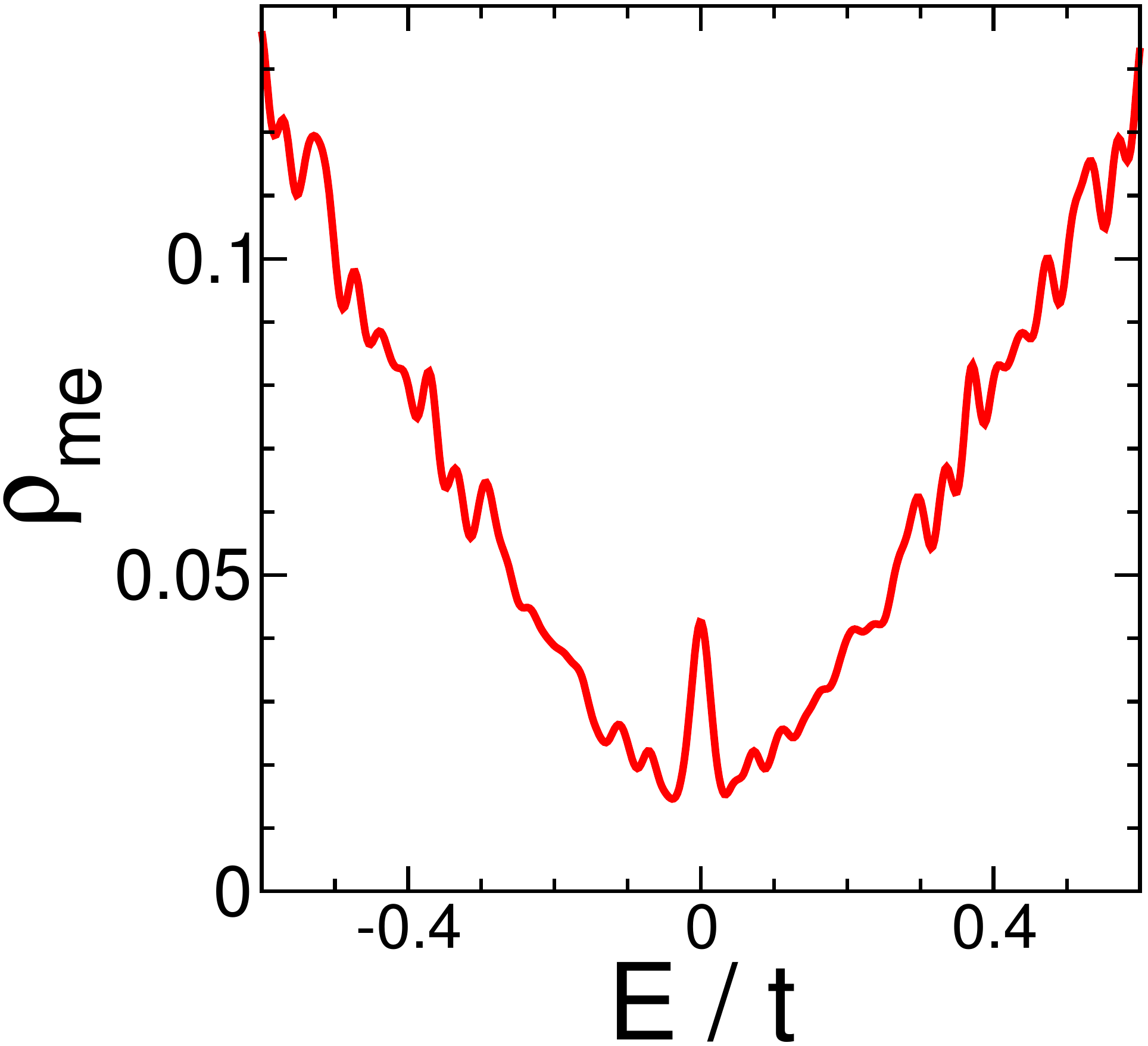}\hspace*{0.6cm}
\includegraphics[width=0.61\linewidth,clip]{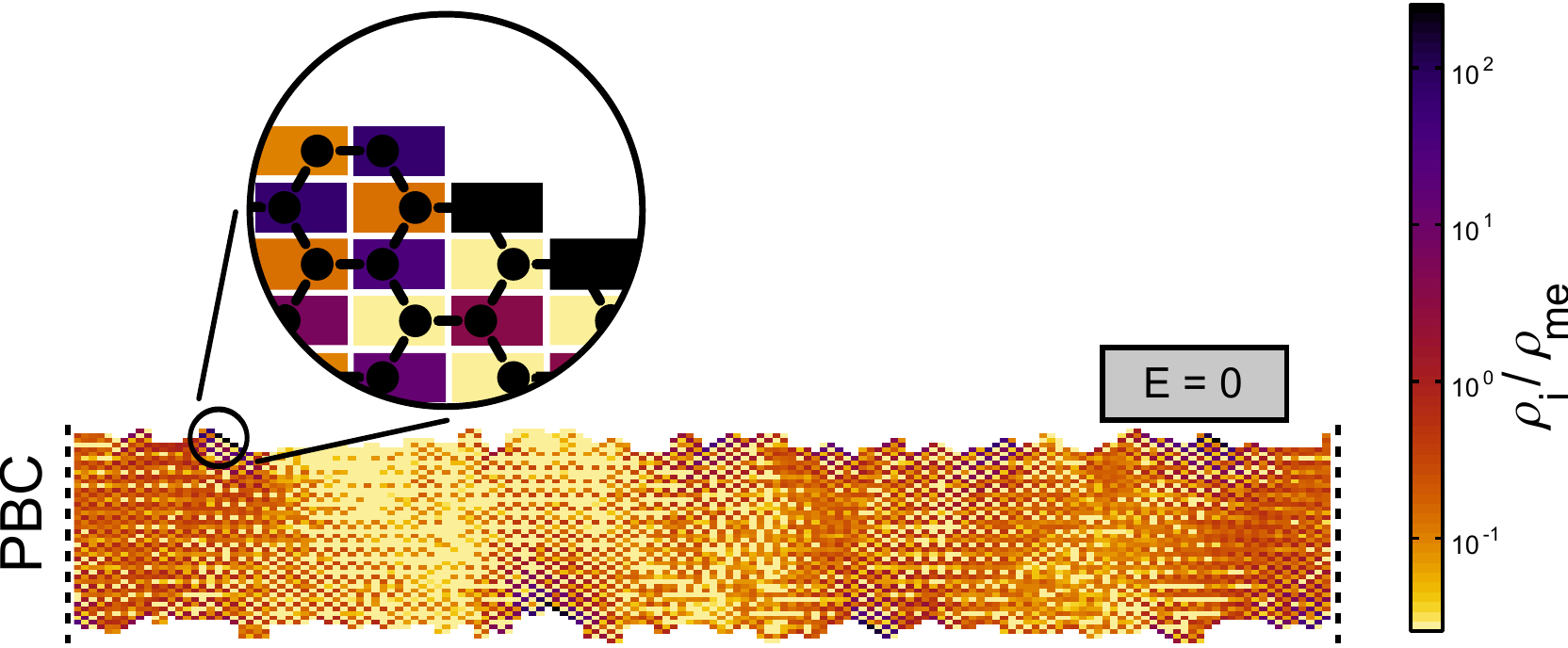}
 \caption{(Color online) Mean DOS and LDOS at $E=0$ for an armchair GNR with edge roughness. The average width of the depicted GNR
 is $\bar{N}_a=N/L_a=41.12$, where $L_a=160$ is the length of the nanoribbon in $x$ direction, where PBC are applied. Here $M=1024$.} 
 \label{fig:rough_surface}
\end{figure*} 

The local properties of a graphene sample with broken translational invariance are best reflected by the local density of states (LDOS),
\begin{equation}
\rho_i(E)=\sum_{n=1}^N |\langle i | n\rangle |^2 \delta (E-E_n)\,,
\label{ldos}
\end{equation}
where $|i\rangle= c_i^\dagger |0\rangle$, and $|n\rangle$ is a single-electron eigenstate of $H$ with energy $E_n$. 
Experimentally the LDOS is directly probed by scanning tunneling microscopy.\cite{NKF09} Theoretically, 
$\rho_i(E)$ can be determined to, {\it de facto}, arbitrary precision by the kernel polynomial method (KPM), which is based on the expansion of the rescaled Hamiltonian into a finite series of Chebyshev polynomials.\cite{WWAF06,WF08} Exploiting the local distribution approach,\cite{AF05,AF08a} the distribution of the LDOS may be used to distinguish localized from extending states, e.g., in order to address the problem of Anderson localization in graphene.\cite{SSBFV10,SF12a} 

\subsubsection{Regular internal boundaries}
In addition to the  extraordinary bulk properties of graphene, finite graphene structures have very interesting surface (edge or boundary) states that do not exist in other systems. For example, the spectrum of GNRs depends on the nature of their edges: zigzag or armchair.\cite{NFDD96,BF06a,Ez06,Ez07} The experimental ability to prepare zigzag edges selectively by an (anisotropic) crystallographic etching process was demonstrated quite recently.\cite{OBHYKSSWE13} A zigzag GNR [with periodic boundary conditions (PBC) along the $x$ direction] presents a band of zero-energy modes. This band is due to surface
states living at and close to the graphene edges. In contrast, the density of states of armchair  GNRs is gapped at $E=0$. We note that zigzag GNRs with hydrogen passivation might also have a gapped band structure provided that edge magnetization exists,\cite{SCL06,GALMW07} which is not very likely at least at room temperatures however.\cite{KOQF11}

Localized states can also appear if a boundary inside the graphene material exists. This is demonstrated by Fig.~\ref{fig:los_void} for a ``regular'' void, realized via infinite on-site potentials $V_i$. The magnifications show that the internal boundaries are of zigzag and armchair types. The four zigzag boundaries give reason to a band of edge states  that shows up by a strong peak in the averaged (mean) DOS, $\rho_{\rm me}(E)=N^{-1} \sum_i \rho_i(E)$, at $E=0$, see left panel.  Note that for such GNRs with voids the localized states located at the sublattice with open bonds   do not allow an analytical solution. The additional peaks in the mean DOS are remainders of the sequence of Van Hove singularities, 
appearing in finite GNRs due to their quasi one-dimensionality.\cite{SSF09} 

\subsubsection{Rough external boundaries}
The fabrication procedure of GNRs usually does not yet allow us to control the boundary of a GNR with atomic precision. Hence the edges of GNRs are disordered as a rule on the atomic length scale, with the result that the transport properties might differ significantly from those of ideal GNRs.\cite{MB09,FN11} To  model a rough graphene boundary, we repeatedly remove edge sites (carbon atoms with only two nearest neighbors) from the GNR,\cite{MCL09} just  by setting the corresponding $V_i=\infty$ with probability $p=1/2$. If we create by chance ``antenna'' (carbon atoms with only one neighbor) or isolated atoms, these will be removed as well. Figure~\ref{fig:rough_surface} gives the mean DOS and LDOS for such a GNR with rough edges. Starting from a GNR with ideal armchair edges along the $x$ direction, the typical sample depicted was obtained after 30 reiterations of the above described procedure.  Both the mean DOS and LDOS signal the existence of localized edge states
which arise because small zigzag regions are generated at the GNR boundary by the cropping process. The LDOS furthermore shows that---caused by the edge roughness---the sites in the bulk with weak (or even vanishing) amplitude of the wavefunction form a filamentary network.
Simultaneously the Van Hove singularities are smeared out as an effect of disorder.

\subsection{Momentum-resolved spectral function}
 The KPM\cite{WWAF06,WF08} can also be used to calculate spectral functions and dynamical correlation functions for disordered GNRs. The influence of disorder on the electronic properties  of graphene and GNRs is of particular interest in the vicinity of the Dirac point. 
Angle resolved photoemission spectroscopy  provides the most direct method to investigate the electronic band structure in this region.\cite{BSSHPAMR10} Quite recently also (small scale rotational) disorder effects have been probed by photoemission measurements  for (epitaxial) graphene (on SiC(0001)).\cite{WBSOKCMISHR13} 
 
 Here we investigate GNRs with short-range Anderson disorder, $V_i\in [-\gamma/2, \gamma/2]$,\cite{SSF09} and determine 
the momentum-resolved single-particle spectral function at zero temperature ($T=0$),
  \begin{equation}\label{eq:A_k}
    {A(\vec k, E)} =  \sum_{n=1}^{N} | \langle n | \psi (\vec k) \rangle |^2 \delta (E - E_n)\,,
  \end{equation}
where $| \psi (\vec k) \rangle = (N)^{-1/2} \sum_i \exp(\imag \vec k \vec r_i) c_i^{\dag} | 0 \rangle$ (note that  $| \psi (\vec k) \rangle$ is not  a Bloch eigenstate of infinite graphene due to its sublattice structure). 

\begin{figure*}[t]
  \centering
  \includegraphics[width=\linewidth,clip]{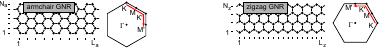}\\
  \includegraphics[width=\linewidth,clip]{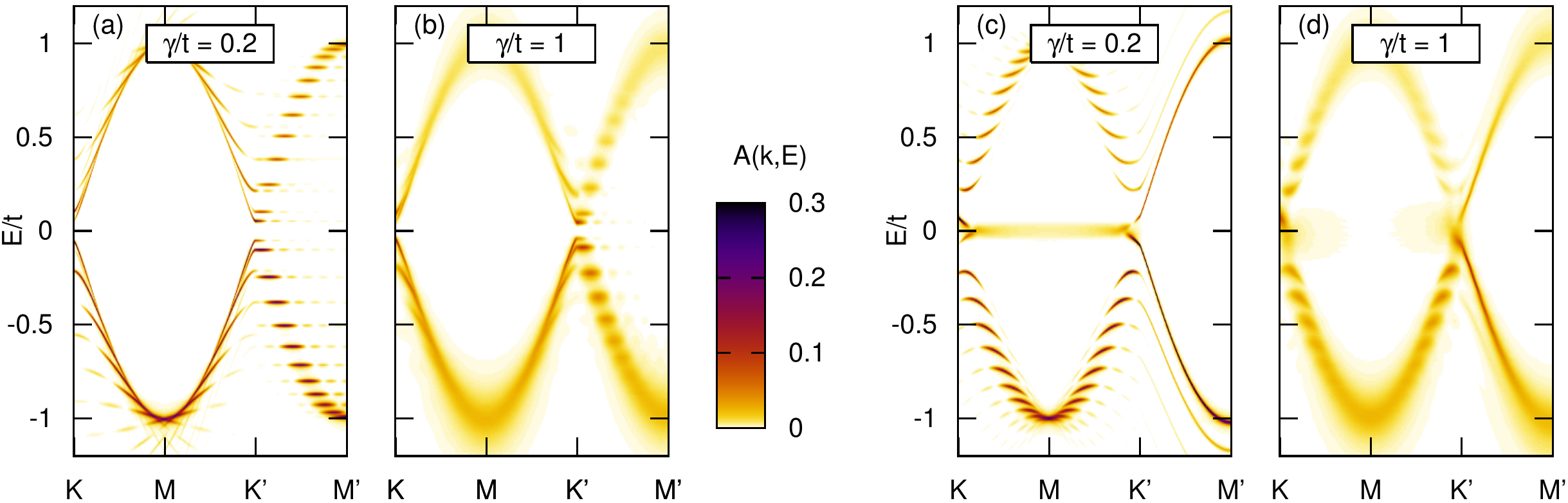}
  \caption{(Color online) 
               Top panels: Armchair and zigzag GNRs  
               and orientation of the corresponding  Brillouin zone (in longitudinal directions PBC are applied).
               Bottom panels: 
                       Averaged spectral function $A(\vec k, E)$ along the red paths indicated in the top panels for Anderson disordered armchair [(a),(b)] and zigzag [(c),(d)] GNRs with  $N_a=34$ (4.18nm), $L_a=1184$ (252nm) and $N_z=20$ (4.26nm), $L_z=2048$ (252nm). 
            Data were sampled over 32 disorder realizations. Within the KPM we use $M=2048$ Chebyshev moments, yielding an energy resolution
            of about $0.005 t$ ($\simeq$ width of the Jackson kernel).                        
                                    }
  \label{fig:spc_function}
\end{figure*}

Figure~\ref{fig:spc_function} presents results for $A(\vec k, E)$ along paths following the Brillouin zone boundary thereby meeting the 
Dirac points $K$ and $K'$. The discreteness of the spectra in the vertical direction is a  finite-size effect (due to the small $N_{a/z}$ in GNRs
with $N=N_{a/z}\times L_{a/z}$ sites), causing a sequence of quasi one-dimensional bands with Van Hove singularities. 
They primarily appear along  the $\overline{K'M'}$ ($\overline{KM}$) direction for armchair (zigzag) GNRs.  
These finite-size signatures will be  readily suppressed by  disorder away from the Dirac points
but persist near $K$, $K'$ even for relatively large values of $\gamma$ [see panel (b)], indicating that the $E\simeq 0$ Dirac fermions  are less affected 
by Anderson disorder. Most notably the almost dispersionsless band of edge states, appearing in zigzag GNRs along  $\overline{KM}$-$\overline{MK'}$ for weak disorder [see panel (c)], is destroyed for strong disorder, where only a few localized edge states reside near the $K$, $K'$ points [see panel (d)].

\begin{figure}[t]
  \centering
   \centering
  \includegraphics[width=0.525\linewidth,clip]{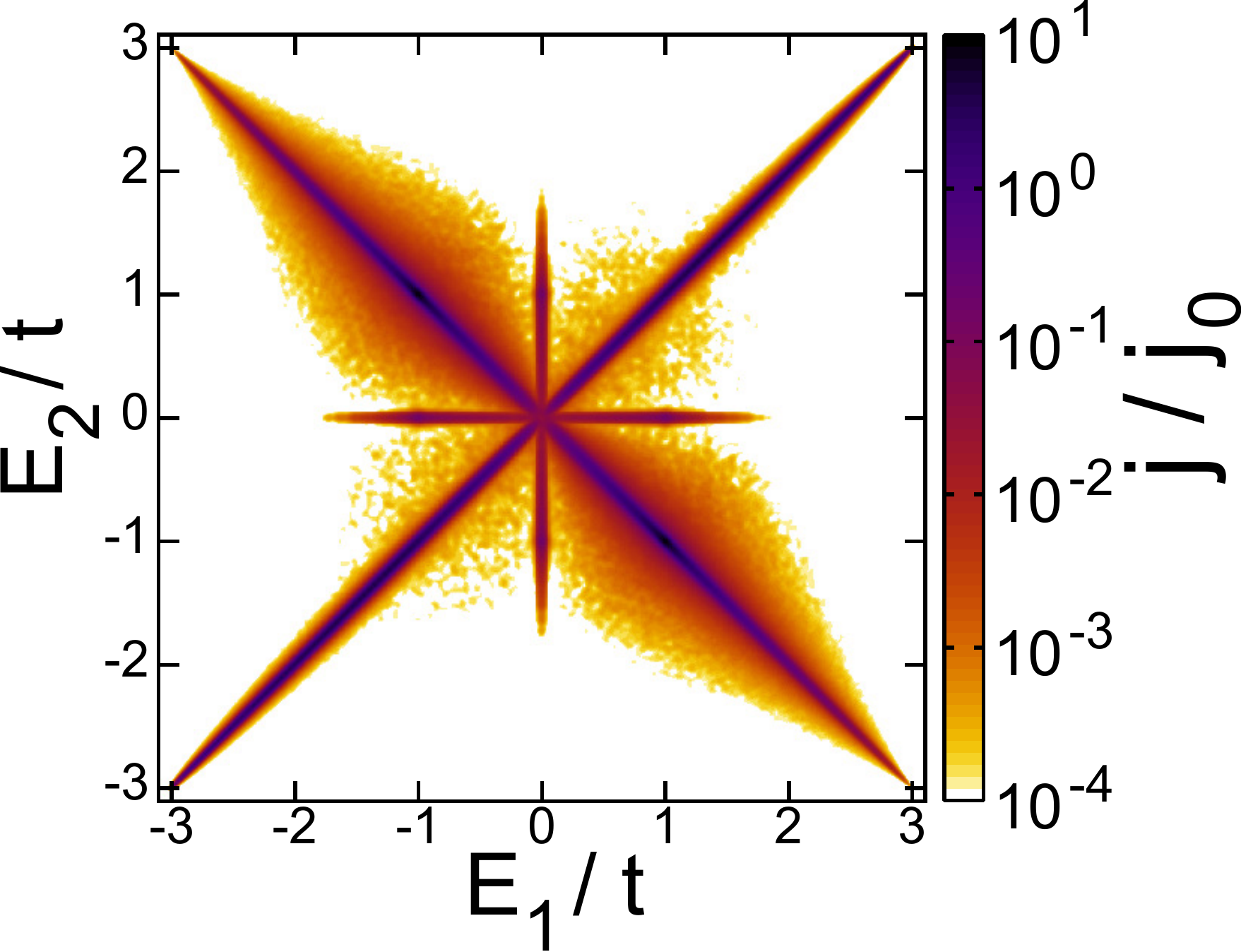}\hspace*{0.1cm}
\includegraphics[width=0.435\linewidth,clip]{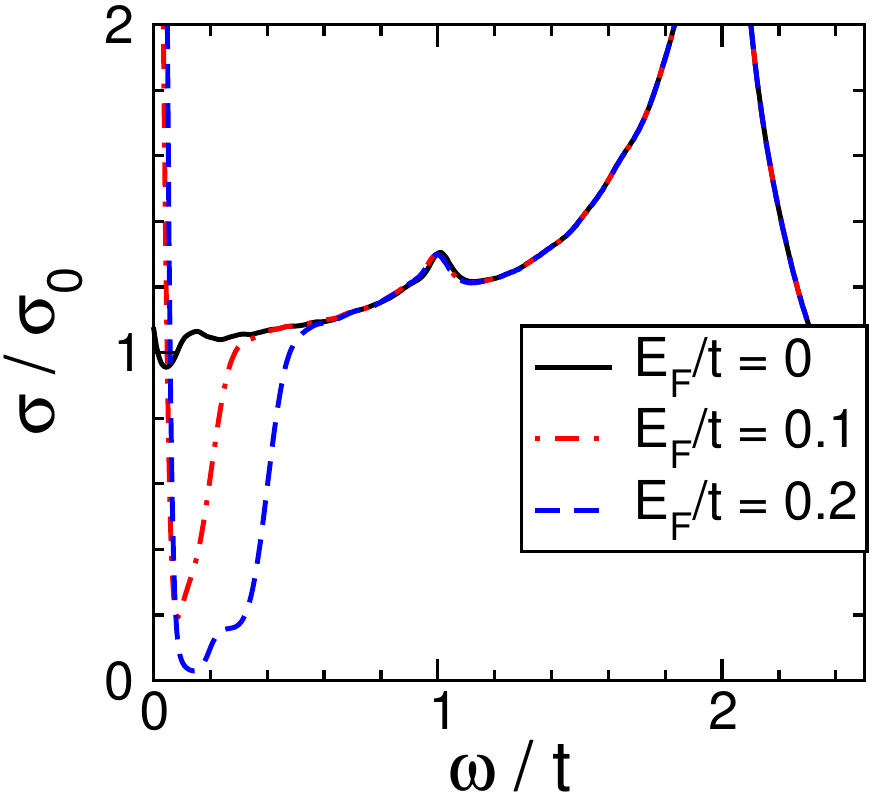}
  \caption{(Color online) 
                 Current matrix-element density $j(E_1,E_2)$ (left panel) and optical conductivity $\sigma(\omega)$ (right panel), where $j_0=e^2/\hbar^2$ and $\sigma_0=e^2/8\hbar$, for a zigzag GNR with $N_z=120$ and PBC in $x$ direction ($L_z=2048$). Calculations of $\sigma(\omega)$ were performed
               at room temperature $T=300$K, i.e., $ \beta = 1/Tk_{\rm B}=108.3/t$. Note that the width of the Jackson kernel used in the KPM  should be smaller than $\beta^{-1}$.
                  }
  \label{fig:cleanZGNR}
\end{figure}

\begin{figure}[hbt]
  \centering
  \includegraphics[width=\linewidth,clip]{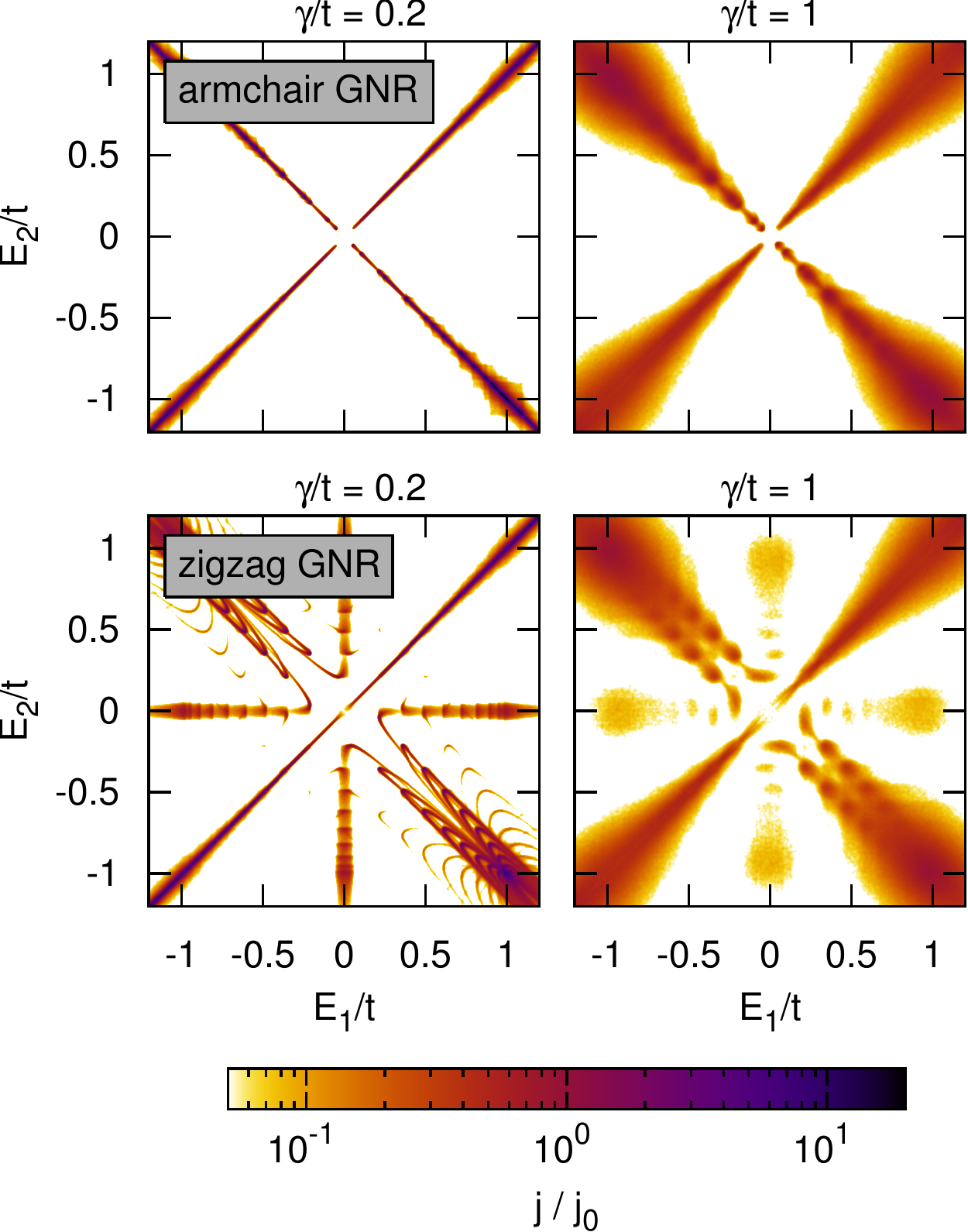}
  \caption{(Color online) 
                 Averaged current matrix-element density $j(E_1,E_2)$ 
                   for  disordered armchair and zigzag GNRs. System parameters are---as in Fig.~\ref{fig:spc_function}---$N_a=34$, $L_a=1184$ (upper panels) and $N_z=20$, $L_z=2048$ (lower panels).
                  }
  \label{fig:j_disorderedgnrs}
  \end{figure}
  \begin{figure*}[t]
  \centering
  \includegraphics[width=.8\linewidth,clip]{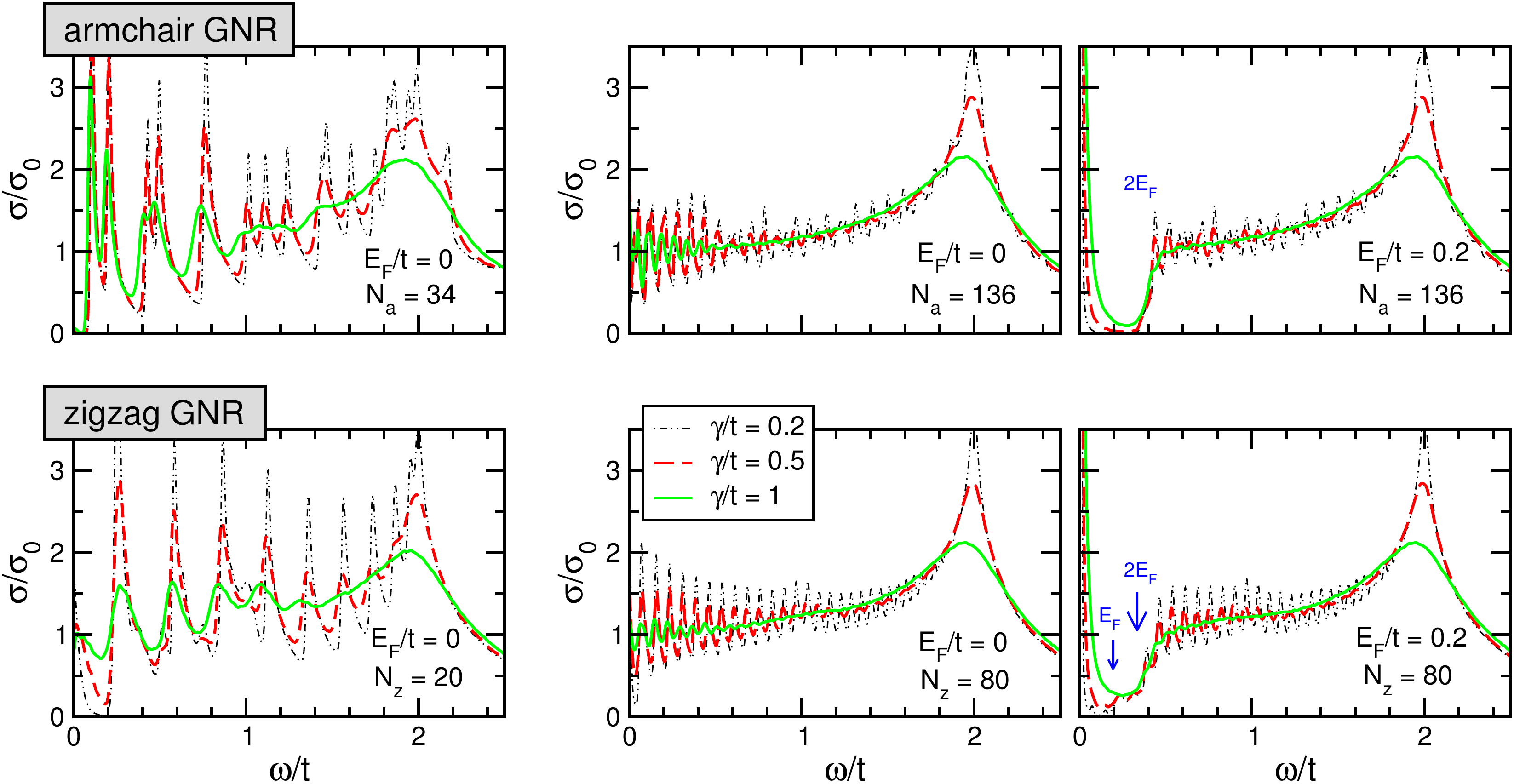}
  \caption{(Color online) 
               Average optical conductivity of disordered armchair and zigzag GNRs at ($E_{\rm F}=0$) and near ($E_{\rm F}/t=0.2$) the charge neutrality point. The ribbon width  is about $4.2\text{nm}$ for the leftmost panels and  $16.8\text{nm}$ for the four right-hand panels. 
               In all cases the ribbon length is about $252\text{nm}$  ($L_a = 1184$, $L_z = 2048$) with PBC in $x$ direction.   }
                
  \label{fig:sigma}
\end{figure*}

\subsection{Optical conductivity}
We next analyze the optical response of disordered GNRs by calculating the so-called regular contribution to the real part of the optical  
conductivity 
  \begin{multline} \label{eq_reg_opt_Leitfahigkeit}
  \sigma(\omega) = 
          \frac{\pi \hbar}{\omega\Omega}\sum_{n,m} |\langle n \mid  J_x \mid m \rangle |^2\,[f(E_n) - f(E_m)]\\\times\delta \left(  \omega + E_n - E_m \right) \\ = \frac {\pi \hbar}{\omega} 
        \int_{-\infty}^{\infty} \text dE  \, j(E, E + \omega) \, \left[ f(E) - f(E+ \omega) \right] 
  \end{multline}
by our KPM scheme.\cite{WWAF06}
In Eq.~\eqref{eq_reg_opt_Leitfahigkeit}, the current operator $J_x=-(\text{ie}t/\hbar)\sum_{\langle i,j\rangle} (x_j-x_i) c_i^\dagger c_j^{} $ ($x_j$ denotes the $x$ component of the position vector $\vec{r}_j$) , the Fermi function  $f(E)= [\E^{\beta (E - E_{\text F})}+1]^{-1}$, and $\Omega=3^{3/2}Na^2/4$ with $a\simeq 1.42$~\r{A}. 

Valuable information about the transport properties can already be obtained from the temperature- and filling-independent
current matrix-element density 
  \begin{equation}
j(E_1,E_2)= \frac{1}{\Omega} {\rm Tr} [J_x \delta (E_1-H)\, J_x \delta (E_2 -H)]\,.
\end{equation}
Here the trace can be evaluated by a stochastic method using a small number---in our case ten---randomly chosen states
for each sample.\cite{SK88,WWAF06} 

For graphene, the current matrix-element density exhibits finite spetral weight only on  an ``X''-shaped support in the $E_1$-$E_2$ plane, where the line $E_1=E_2$ accounts for the dc conductivity ($\omega=0$). The line  $E_1=-E_2$, on the other hand, describes the ac optical response due to vertical $\pi$-$\pi^\ast$ interband transitions (recall that $H$ and $J_x$ do not commute for the non-interacting graphene honeycomb lattice model).  For GNRs boundary effects will strongly affect these signatures. 

Figure ~\ref{fig:cleanZGNR} displays $j(E_1,E_2)$ and  $\sigma(\omega)$ for the zigzag case.  First of all the spectral signature at $E_1=-E_2$ widens out. Of higher significance, however,  will be the additional ``$+$''-shaped absorption feature which can be attributed to optical transitions between edge and bulk states. The optical conductivity at 
fixed $\omega$, according to the second line of Eq.~\eqref{eq_reg_opt_Leitfahigkeit}, 
is given by  an integral over $j(E_1=E,E_2=E+\omega)$,  where the Fermi factors filter
out  contributions located in the second and fourth quadrant only.  Furthermore, they suppress $\sigma(\omega)$
below $\omega=2|E_{\rm F}|$, yielding a step feature.  If compared to the optical response of bulk graphene, 
showing besides this step a single maximum at  the DOS Van Hove singularity point  $\omega=2t$ only, the 
edge states in zigzag GNRs lead to  a further step at $\omega=|E_F|$ and an additional local maximum at $\omega=t$, see right panel. 
Clearly, for $\omega\to 0$, we find a Drude peak at $|E_F|>0$, i.e. at finite filling, whereas $\sigma\to\sigma_0$ at the charge neutrality point ($E_F=0$).

Figure~\ref{fig:j_disorderedgnrs} contrasts the current density for disordered armchair and zigzag GNRs for the same parameters as the spectral function was shown in Fig~\ref{fig:spc_function}. In armchair GNRs the spectral weight is appreciable near  the ``X''-shaped support for the regular and weakly disorders cases. Near the origin $E_1=E_2=0$ disorder effects are almost  negligible. At large 
$\gamma$ a broading sets in that emanates from the point $|E_{1,2}|=\pm t$.  This effect is also observed for zigzag GNRs but superimposed by the ``$+$'' shaped absorption feature, which becomes broadened by the disorder as well.  Since the zigzag GNR under consideration has a noticeable smaller width compared with that used in  Fig.~\ref{fig:cleanZGNR}, finite-size effects influence the optical transitions in the vicinity of $E_1=-E_2$ even for relatively strong disorder, particularly when  $|E_{1,2}|\leq 0.5 t$.

The resulting optical response of armchair and zigzag GNRs of different width is shown in Fig.~\ref{fig:sigma}  
for different disorder strengths $\gamma$. Obviously the conductivity of narrow GNRs is dominated by Van Hove DOS effects. Naturally the corresponding maxima in $\sigma(\omega)$ weaken as the ribbon width increases, except the one at $\omega=2t$ caused by the 2D graphene honeycomb lattice structure. The higher the optical frequency,  the stronger these finite-size signatures will be reduced by Anderson disorder.  Also the maximum at $\omega=2t$ is suppressed. On the other hand, away from the charge neutrality point, 
disorder enhances the optical response in the region $0< \omega \leq 2|E_{\rm F}|$, because the (Pauli) blocking of states could 
be locally  overcome. Furthermore, we find that the above mentioned fingerprints of edge states in zigzag GNRs survive the presence of disorder 
to a large extent. 

To subsume, the combined LDOS, single-particle spectral function and optical conductivity results presented in this section 
give a largely consistent picture of how disorder affects the electronic and transport properties of isolated GNRs.

\section{Normal-conductor graphene junctions}
\begin{figure}[h]
  \centering
  \includegraphics[width=0.6\linewidth,clip]{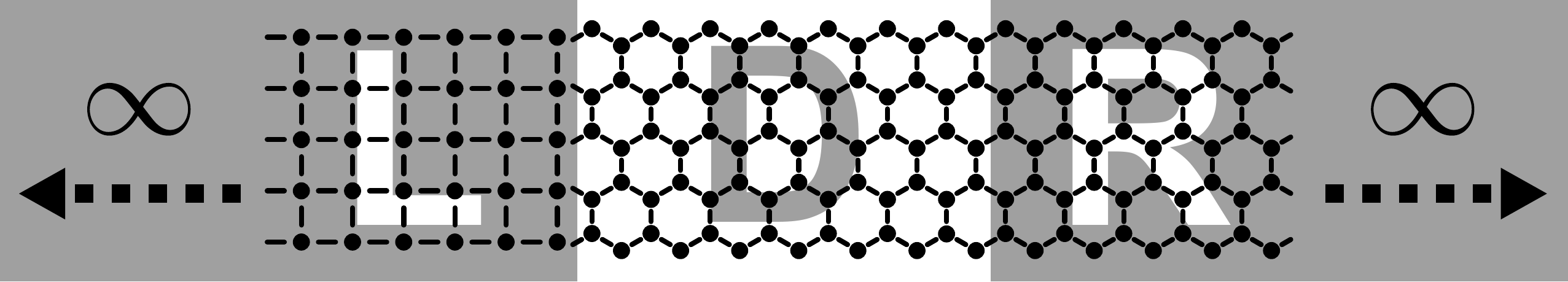}
  \caption{Schematics of an end-contacted zigzag GNR.  
            The leads L and R are regular infinite quantum wires; here the left (right) wire is asumed to be a metal (graphene).  
            Only the graphene device part  D might be influenced by disorder. The left (right) interface forms a NG (GG) junction.
                        }
  \label{fig:GFT_demo}
\end{figure}

In a next step, we address the (linear) transport through disordered GNR in the lead-sample geometry most relevant for experiments. 
The ultimate electronic contacts are metallic or gated graphene. The coupling between graphene and the metallic electrodes can be realized by hybridization. In general it is extremely difficult  to describe this coupling within {\it ab initio} approaches. Therefore simplified models have been proposed to describe normal-metal/graphene (NG) junctions.\cite{BM07,RS07,DD10,ZQ10,ZGZLQS12} We consider here only two-terminal contact systems, as schematized in Fig.~\ref{fig:GFT_demo}.  End-contacts, instead of side-contacted setups, might be justified when the nanostructure/electrode coupling is strong, which means good transparency of the interfaces and weak chemical bonding.\cite{Kr12}

For small bias voltages, within the Landauer-B\"uttiker formalism, the phase-coherent conductance $G(E)/G_0$  equals the transmission function\cite{Dat95,FG97}
\begin{equation}
T(E)=\trace [\Gamma_{\text L}            \greenf_{\text D} 
                                                            \Gamma_{\text R}  
                                                            {\greenf_{\text D}}^\dagger]|_E\,,
\end{equation}
where the lead-system coupling matrices 
\begin{equation}
\Gamma_ {\text L/R}(E)=\text{i}[\Sigma_{\text L/R}-\Sigma^\dagger_{\text L/R}]|_E
\end{equation}
contain the self-energies  $\Sigma_{\text L/R}$ describing the contacts (L/R), and the Green function of the system (D) is
\begin{equation}
\greenf_{\text D}  (E)=\frac{1}{E-{H}_D-\Sigma_{\text L}-\Sigma_{\text R}}\,.
\end{equation}
In case of a homogeneous ballistic conductor, $T(E)$ simply counts the number of propagating modes. Each transport channel contributes $G_0 = {\text e}^2/h$ to the conductance. For a zigzag
GNR we have $T(0)=1$. $T(E)$ then increases away from the charge neutrality point (for large $N_z$ almost straight proportional to $E$) and reaches its maximum  $T^{\rm max}=N_z$ at $E=\pm t$, before it falls off again and becomes zero at the edge of the spectrum $T(\pm 3t)=0$.  Modeling the metallic lead by a tight-binding square-lattice Hamiltonian with $N_m$ sites in $y$ direction, we find $T(0)=T^{\rm max}=N_m$, $T(\pm 4t)=0$, and 
$T(E)=N_m \pi^{-1} \arccos(|E|/(2t)-1)$ as $N_m\to \infty$. 
The NG interface couples these modes to the propagating and evanescent modes in the graphene scattering region D.

\subsection{Influence of contacts}
\begin{figure}[t]
  \centering
  \includegraphics[width=0.24\linewidth,clip]{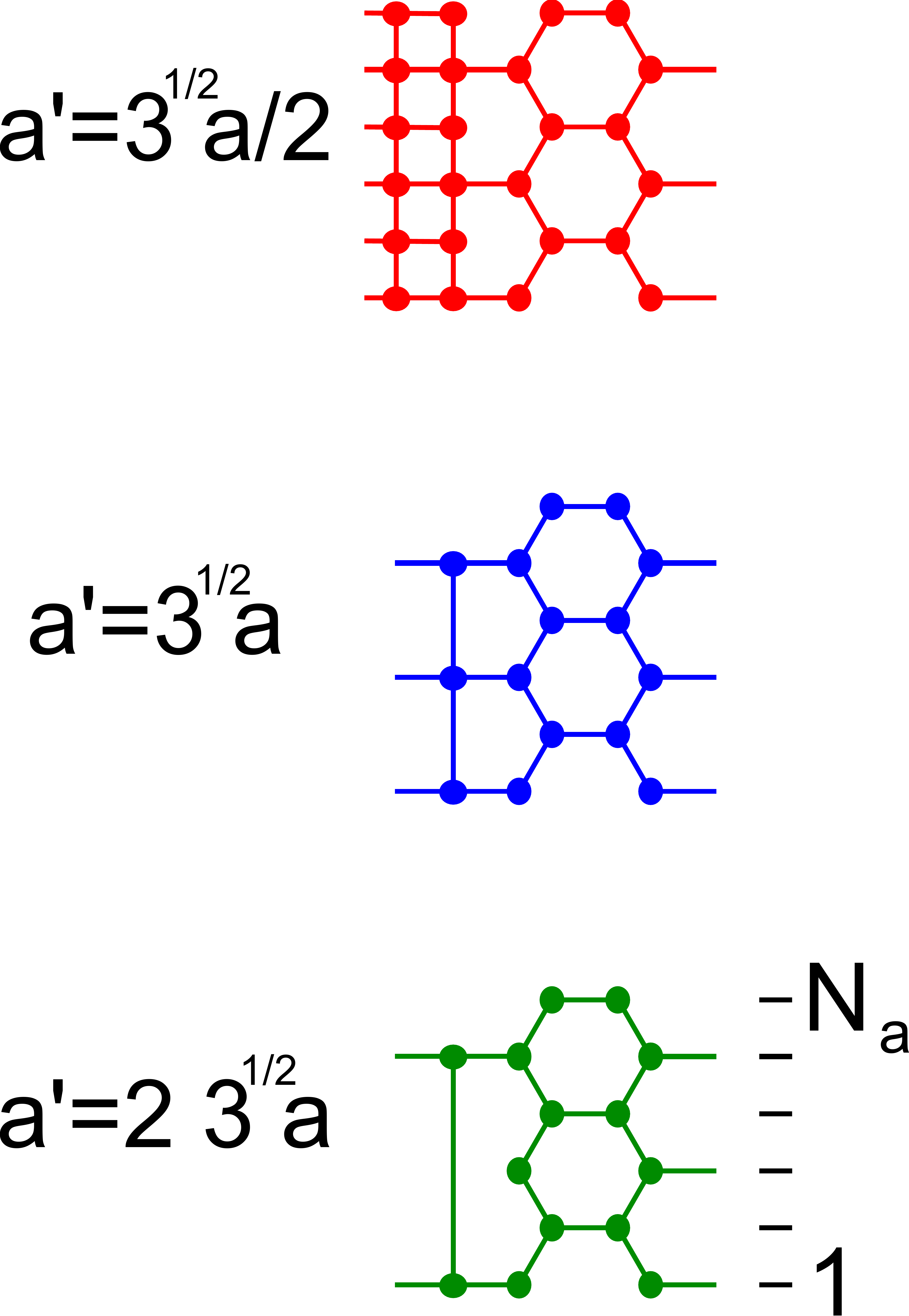}
  \hspace{0.05cm}
  \includegraphics[width=0.68\linewidth,clip]{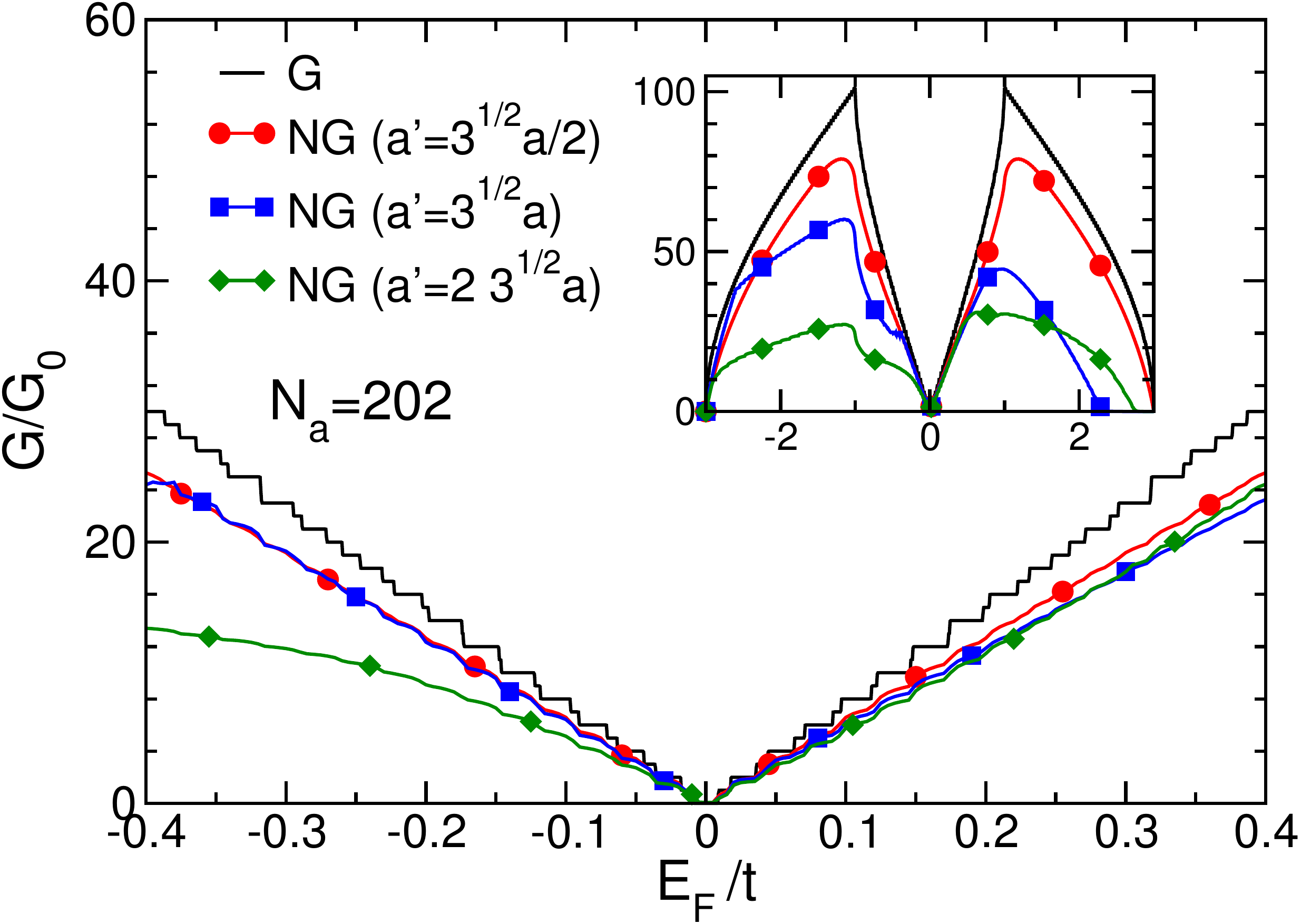}
  \caption{(Color online) Conductance of an NG junction with zigzag interface of armchair GNR contacted to different real-space square-lattice leads.  The N (G) ribbon is semi-infinite to the left (right). To compare with, the conductance of a pure GNR is included, see black solid line.     The main panel displays the regime close to the charge neutrality point, whereas the inset gives  $G/G_0$  for  the whole energy band. }
  \label{fig:acMG}
\end{figure}

\begin{figure}[t]
  \centering
  \includegraphics[width=0.24\linewidth,clip]{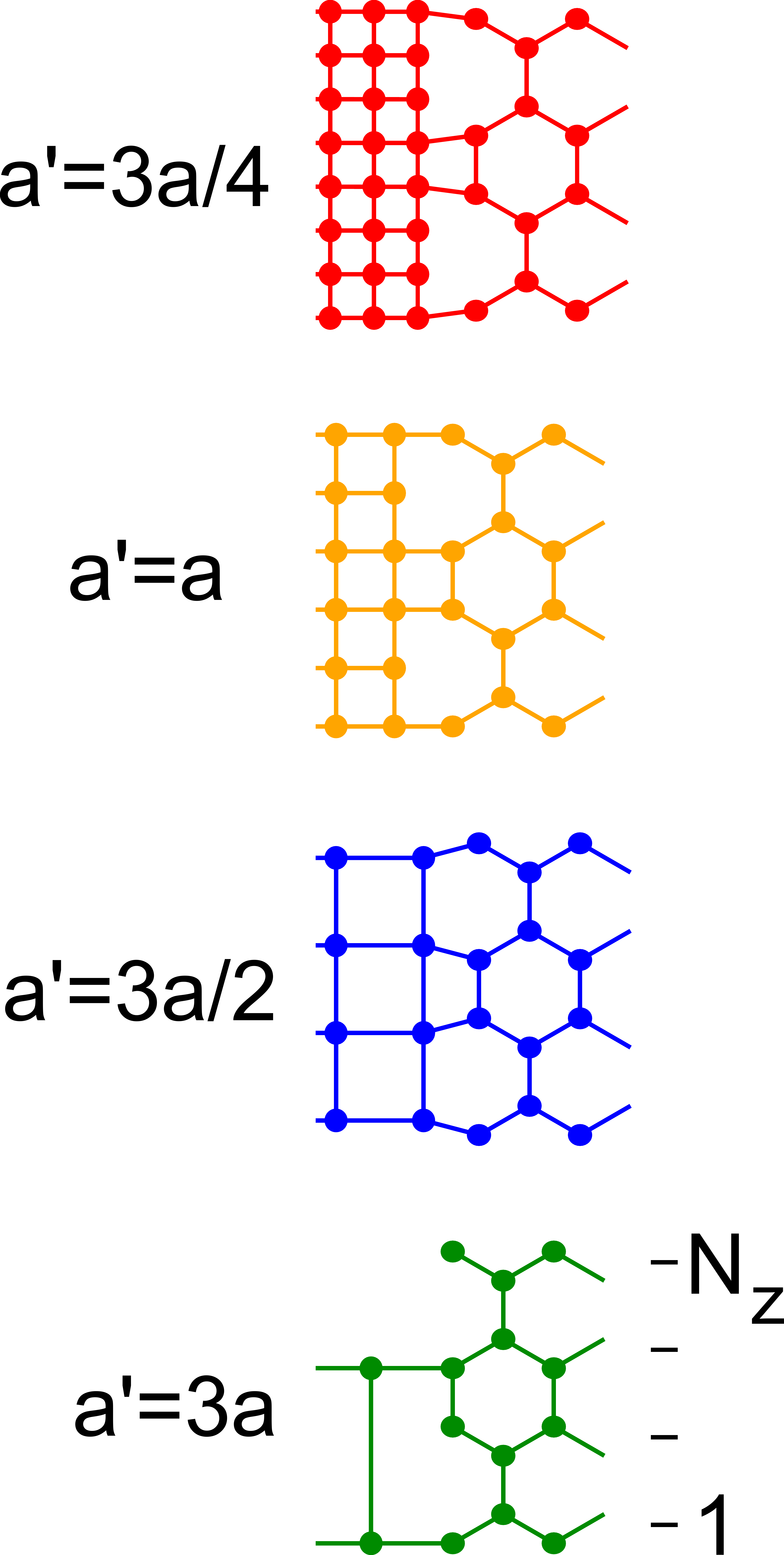}
  \hspace{0.05cm}
  \includegraphics[width=0.68\linewidth,clip]{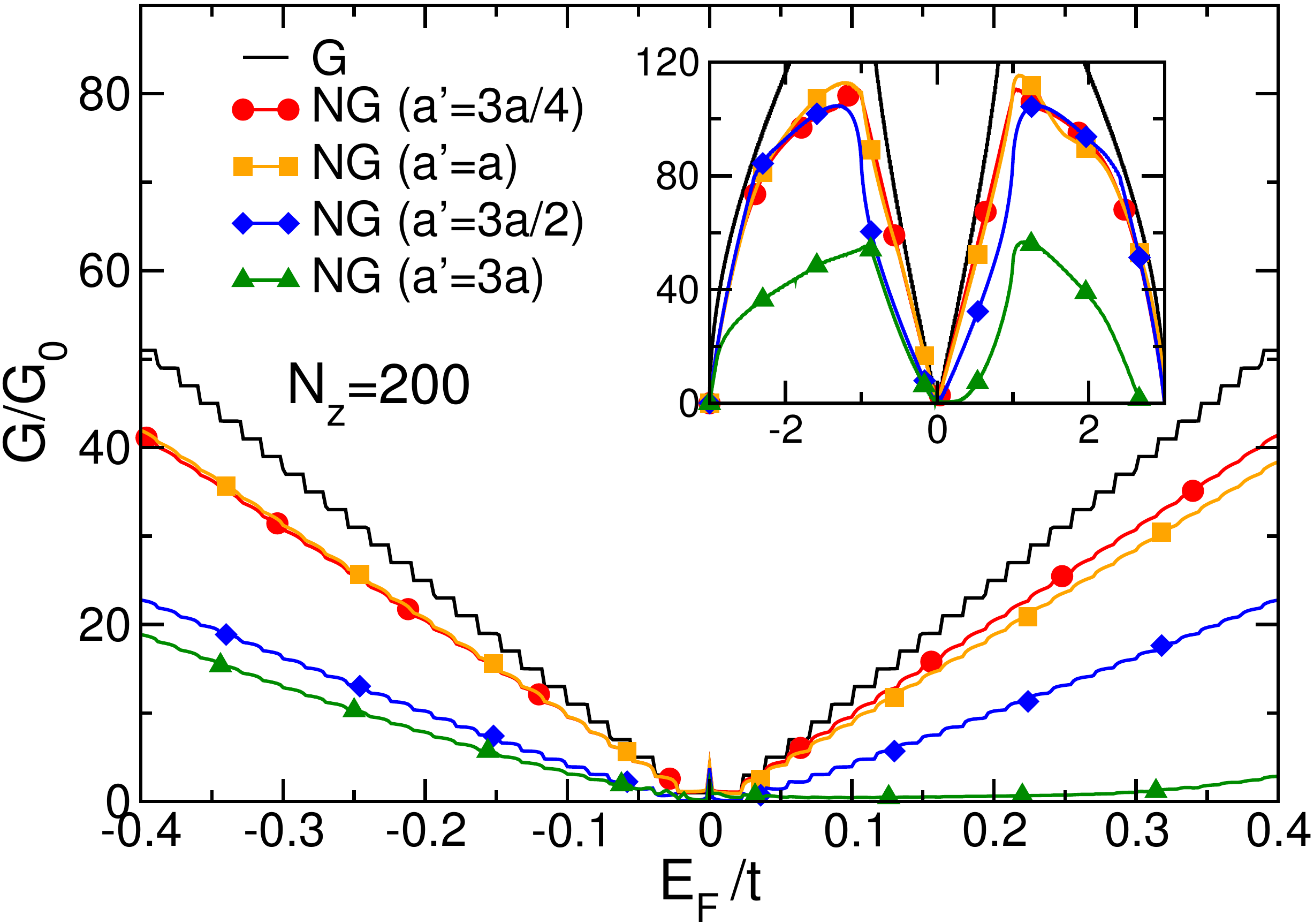}
  %Idee: Inlet mit Vergroesserung auf E=0
  \caption{(Color online) Conductance of an NG junction with armchair  interface of zigzag  GNR contacted to different real-space square-lattice leads.    The setup is the same as in Fig.~\ref{fig:acMG}. }
  \label{fig:zzMG}
\end{figure}

Let us first consider zigzag (armchair) graphene interfaces connected to square-lattice-matched metallic leads  by
assuming  that the lattice constant of the metal $a'=\{3a/4\,,  a\,, 3a/2\,, 3a \}$ ($a'=\{\sqrt{3/4}a\,,  \sqrt{3}a\,, 2\sqrt{3}a\}$). Furthermore, we fix
$a' N_m \simeq 1.5 a N_z$ ($a' N_m \simeq  \sqrt{3/4} a N_a$)  and hypothesize the same 
Fermi energy $E_{\rm F}$ and transfer amplitude $t$ to exist in the metal respectively NG link.  Then conductance shows the behavior displayed in Fig.~\ref{fig:acMG} (armchair GNR with metal-to-zigzag-graphene interface) and   Fig.~\ref{fig:zzMG} (zigzag GNR with metal to armchair-graphene interface). Most notably, close to the charge neutrality point, we observe in all cases an almost linear dependency 
of the conductance on the absolute value of the Fermi energy $|E_{\rm F}|$. The slope in the $p$- ($E_{\rm F}<0$) and $n$-type ($E_{\rm F}>0$) regime in general differs however.  Armchair GNRs with zigzag interface match better to metallic quantum wires because there exists an equivalent mode selection.\cite{Sch07} Obviously  the slope near $E_{\rm F}=0$ does not much depend on $a$'.  We nevertheless observe a significant overall reduction of $G(E)$ and a strong asymmetry for $E_{\rm F}\to - E_{\rm F}$ away from the Dirac point,  if there are dangling bonds on the GNR side of the interface.  The same holds for zigzag GNRs with armchair interfaces but here the latter effect leads to a stronger reduction of $G$ in the $n$-type regime.


\begin{figure}[t]
  \centering
  \includegraphics[width=\linewidth,clip]{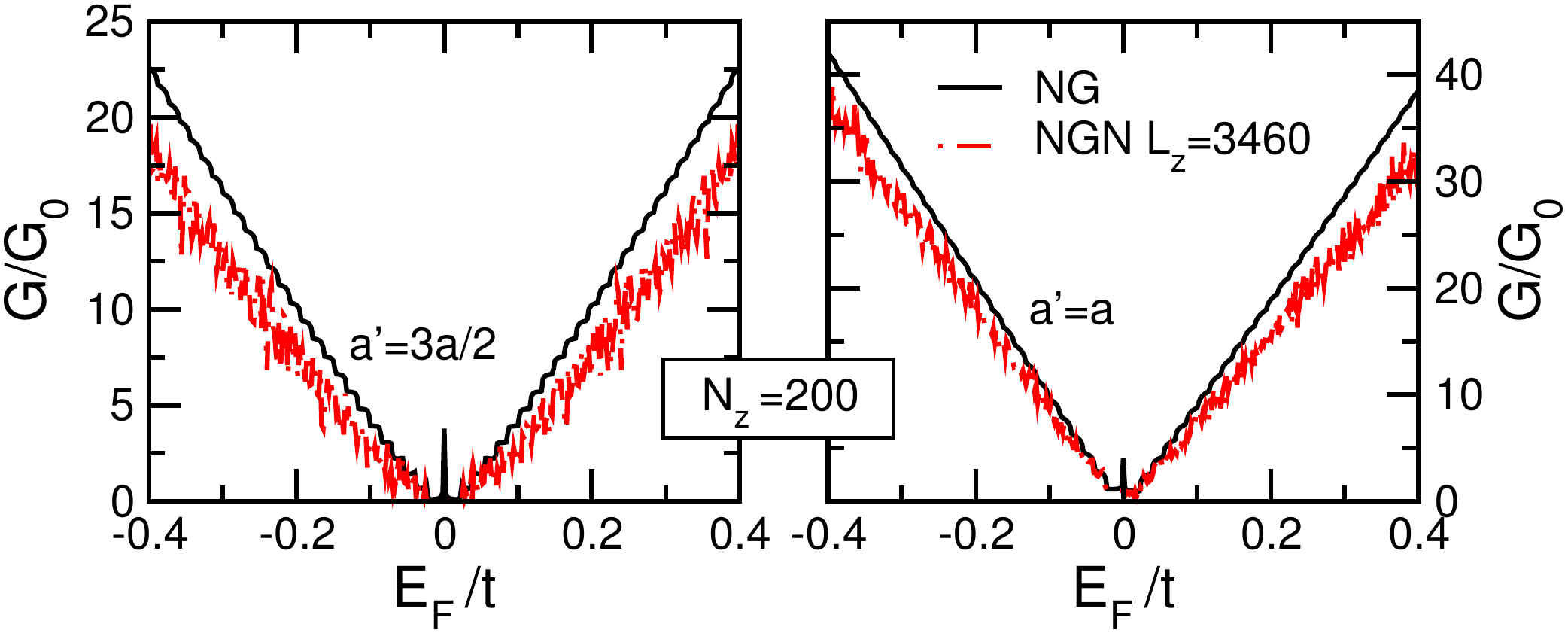}
  \caption{(Color online) 
               Conductance of  an (NGN) junction with semi-infinite metallic leads
               embedding a zigzag GNR with $N_z=200$, $L_z=3460$.  
               The armchair interface contacts used have $a'=3a/2$, allowing for a symmetric mode matching (left panel), 
               and $a'=a$ with dangling bonds on the metallic side of the interface (right panel); c.f. Fig. \ref{fig:zzMG}.
               The black solid curve includes the corresponding data  for a single NG junction.
               }
  \label{fig:clean_NGN}
\end{figure}

\begin{figure}[t]
  \centering
  \includegraphics[width=\linewidth,clip]{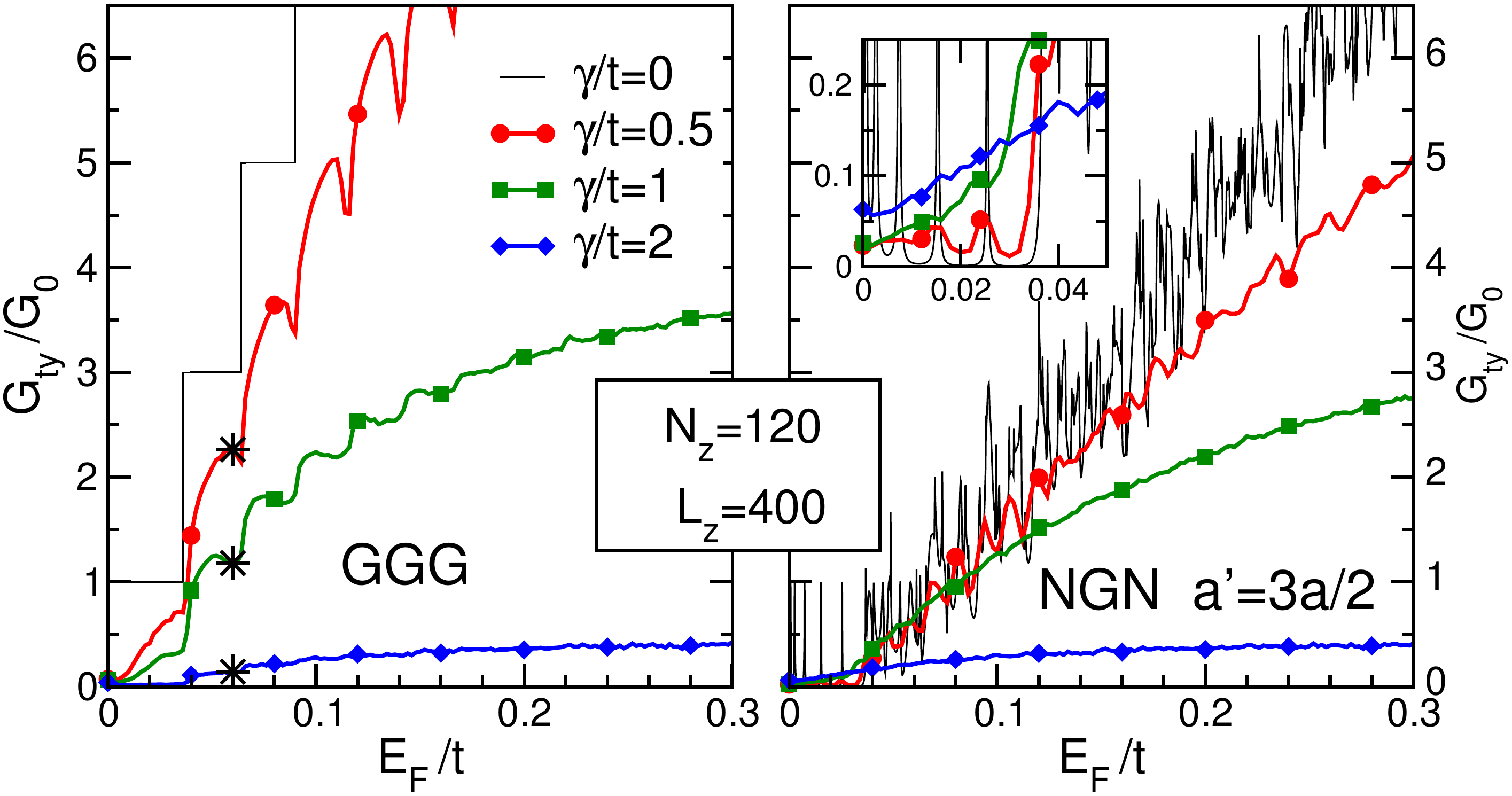}
  \includegraphics[width=\linewidth,clip]{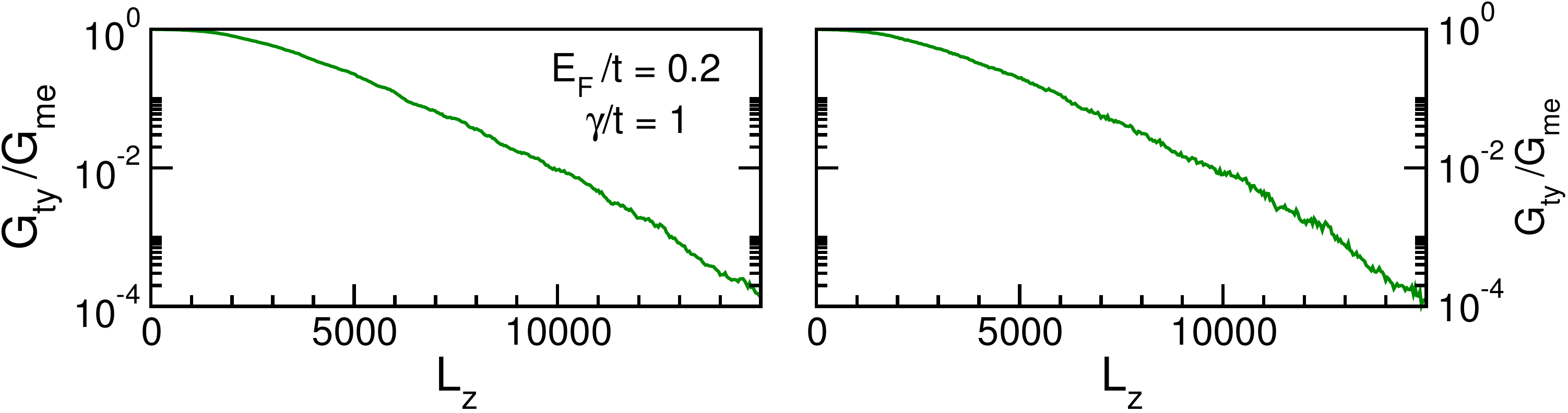}
  \caption{(Color online) Typical conductance of GGG (left) and NGN (right) junctions as a function of the Fermi energy (upper panels), 
  where the central graphene device part is built up of Anderson disordered zigzag GNR.
   $S=1000$ realizations are used for averaging. For comparison data for clean junctions included by black lines.  
   Note that for the configurations marked by stars the distribution of the conductances is given in Fig.~\ref{fig:GGG_distribution}. 
   In the lower panels we show $G_{\rm ty}/G_{\rm me}$ for GGG and NGN setups in dependence on the GNR length at fixed energy $E_{\rm F}/t=0.2$ and disorder strength $\gamma/t=1$. 
 }              
                     \label{fig:MGM_disorder}
\end{figure}

\begin{figure}[t]
  \centering
  \includegraphics[width=\linewidth,clip]{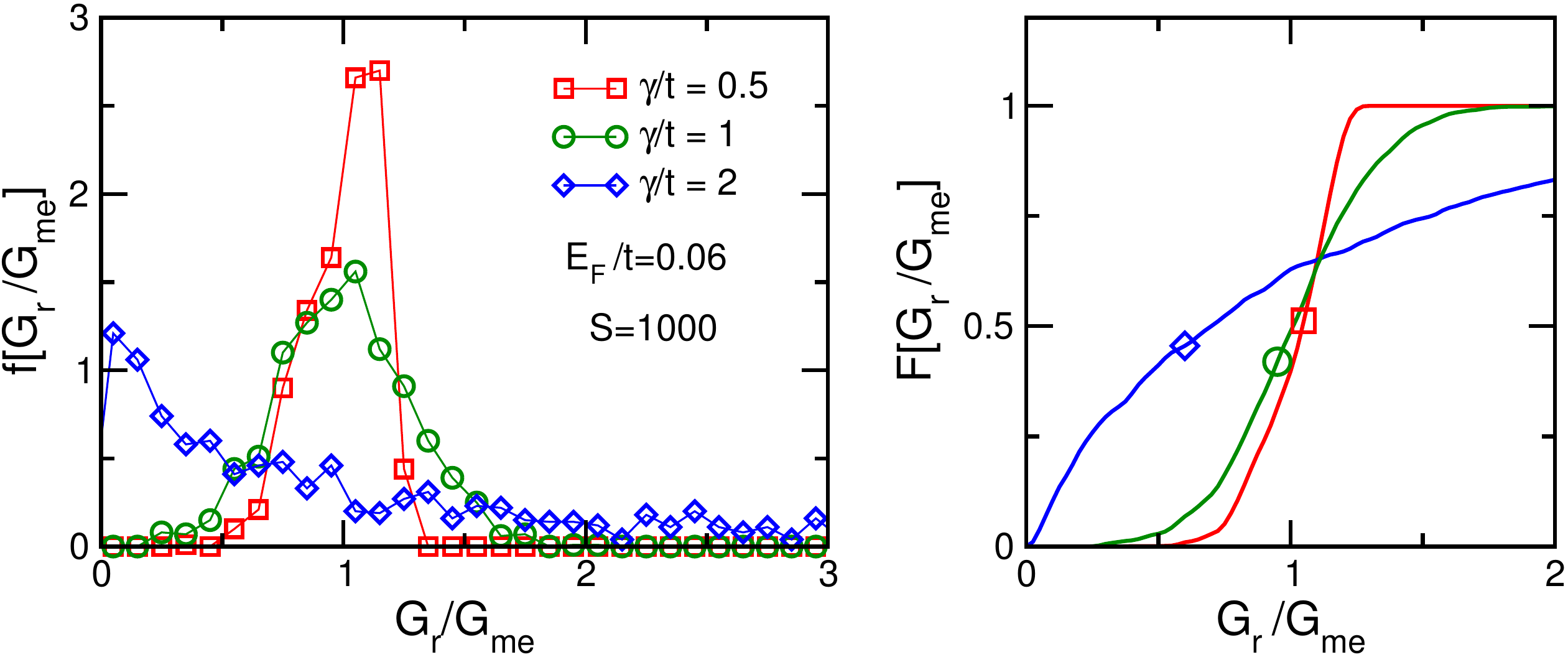}
  \caption{(Color online) Normalized conductance distribution function $f(G_r/G_{\rm me})$ (left panel) and integrated distribution function $F(G_r/G_{\rm me})$   (right panel) for disordered GGG  with $N_z=120$ and  $L_z=400$, as obtained---at $E=0.06t$---for different disorder strengths $\gamma$ from $S=1000$ samples. }
  \label{fig:GGG_distribution}
\end{figure}

Figure~\ref{fig:clean_NGN} shows the conductance for a clean normal-metal /graphene/normal-metal (NGN) junction, where a zigzag GNR with a length-to-width aspect ratio of ten was sandwiched between two metallic leads. Since the intervening GNR acts as a scattering center, $G$ is smaller than for a  both-sided semi-infinite NG junction. The scattering gives rise to a strongly fluctuating $G$.   The average value of $G$ is higher for the interface with $a'=a$ because there are more transport channels (modes) available on the metal sides of the contacts.  

\subsection{Effects of impurity scattering}
Finally we investigate disordered GNRs encased by ordered graphene or metal leads. Defects and impurities are inevitable in graphene based devices. Anderson disorder---with on-site  potentials $V_i$  drawn from a box distribution---can be used to model the effects of short-range impurity scattering by local imperfections. Quite generally the treatment of disordered systems requires the study of distributions of physical quantities, making the application of statistical methods necessary.\cite{ATA73,AF05} In particular this applies to the investigation of subtle disorder phenomena, such as Anderson localization.\cite{SSBFV10} To characterize  the transport through disordered graphene junctions the conductance should be analyzed in this manner. 

Calculating the conductance for $S$ realizations  of an end-contacted disordered (zigzag) GNR (see Fig.~\ref{fig:GFT_demo}),
the mean sample-averaged conductance  $G_{\text {me}}   =  \tfrac 1 S  \sum_{r=1}^S  G_r$  
strongly fluctuates if $G_{\text {me}}/G_0$ gets small. This particularly happens near the Dirac point, for large disorder and long
GNRs, i.e., when Anderson localization induced states appear.  Then the conductance, just as the LDOS,\cite{SSBFV10}  exhibits a log-normal
(rather than a normal) distribution, whose maximum is strongly finite-size dependent. If $L_z\to\infty$ the probability distribution $f[G_r(E)/G_{\text {me}}]$ becomes even  singular at $G_r/G_{\text {me}}=0$. To reflect  such  behavior  
the typical  conductance,        
\begin{equation} %\label{}
     G_{\text {ty}} (E)  = \exp \left( \frac 1 S \sum_{r=1}^S \ln G_r  (E) \right) \, ,
\end{equation}
is more appropriate than $G_{\text {me}}$ ({\it inter alia} because it puts sufficient weight on small values of $G_r(E)$).

Figure~\ref{fig:MGM_disorder} gives the variation of $G_{\text {ty}}$ with the (Fermi) energy for different disorder strengths. 
Here the left (right) panel shows a GGG (NGN) junction. Without disorder, the GGG realizes a homogeneous GNR and we observe 
the step-growth of $G/G_0$ as $E$ increases, when more and more transport channels become  available. For NGN junctions, a 
fluctuating $G/G_0$ results (see the discussion of Fig.~\ref{fig:clean_NGN}). With disorder, for GGG junctions, an overall reduction 
of $G_{\text {ty}}/G_0$ takes place. Thereby  the steps will be blurred. At very large $\gamma$,  $G_{\text {ty}}/G_0\to 0$
even for finite $L_z$, indicating Anderson localization.  Already for $\gamma\sim 2t$, the transport behavior of 
GGG and NGN junctions is essentially the same. Disorder suppresses the inherent conductance fluctuations of clean NGN junctions, 
thereby it might even enhance the conductance at small $E_{\rm F}$ (see inset).
Interestingly, for weak disorder, we observe a ``negative differential conductance'' in the vicinity of the steps. This effect is more 
pronounced at larger energies and can be ascribed to the small curvature (flatness) of the bands if leaving/entering an old/new transport  
channel, making these states very susceptible to disorder. The lower panels impressively demonstrate that 
$G_{\rm ty}\to 0$  as $L_{z}$ increases at fixed $|E_{\rm F}|>0$  for (any)  $\gamma/t >0$, whereas $G_{\rm me}$ stays finite. 
For $\gamma/t\to 0$, Anderson localization occurs not before $L_z\to\infty$.

We now determine the probability distribution of the conductance, $f[G_{r}/G_0]$, by accumulating the values of $G_r$, calculated at a given energy $E$ for $S$ random samples of fixed size, within a histogram. Out of it the integrated distribution 
\begin{equation}
F[G_r /G_{\rm me}]=\int_0^{G_r} f[G^\prime_r/G_{\rm me}] \,{\rm d}G^\prime_r
\end{equation}
 can be obtained. The results are shown in Fig.~\ref{fig:GGG_distribution}. For weak disorder,
when the finite disordered GNR junctions is transmissible,  the probability distribution of the conductance is rather symmetric and peaked close to its mean value. In this regime the distribution is almost unaffected upon increasing the size of the GNR (not shown). For strong disorder, the distribution of $G_r$ is asymmetric and markedly depends on $L_z$. Although most of the weight now is concentrated close to zero, the distribution extends to very large values of $G_r$. That is, the mean value is much large than the most probable value and does not give any valuable information about the transport behavior.   While for weakly disordered GGG junctions the more or less uniform conductances lead to a steep rise of  $F[G_r /G_{\rm me}]$ at $G_r/G_{\rm me}=1$, for strong disorder a very gradual increase is observed. Disordered NGN junctions show the same tendencies.

\section{Conclusions}
By exact numerics we have studied the electronic and transport properties of finite (but actual sized) graphene based structures and 
demonstrated that the geometry of the sample and in particular edges, voids, contacts, and disorder strongly affects the local 
density of states,  the single-particle spectrum, the ac conductivity  and the conductance, probed scanning tunneling microscopy, angle resolved photoemission spectroscopy, optical response and two-terminal transport measurements, respectively.

We showed that  localized edge states dominate the mean density of states (DOS) of graphene nanoribbons (GNRs)---which feature voids or rough surfaces---near the charge neutrality point. In the latter case,  sites in the edge region having vanishing amplitude  entail a filamentary network of the local DOS in the bulk. For disordered GNRs, both the averaged single-particle spectral function and optical
conductivity indicate that disorder tends to suppress the finite-size effects caused by  the geometry of the nanoribbon. 
The states near the  $K$, $K'$ point are robust against disorder to the greatest possible extent. This does not  apply to the 
band of localized edge states. 

The conductance of edge-contacted GNR sensitively depends on the lead-GNR
matching conditions. In this respect armchair GNRs enable a somewhat better current injection. Dangling bonds on the GNR side of the interface substantially reduce the conductance. The typical conductance of disordered GNRs sandwiched between graphene leads
in a junction setup exhibits a negative differential conductivity whenever new transport channels become available by increasing the
Fermi energy.   This accentuates the efficiency of Anderson localization effects at the band edges of electronically low-dimensional 
systems. For GNR junctions, the conductance distribution function manifests  a precursor of the transition from 
a current-carrying to an (Anderson disorder induced) insulating behavior, which is expected to takes place 
when the size of the  disordered active graphene region becomes infinite.

Finally, let us emphasize once more that all these results were obtained within the limits of a non-interacting nearest-neighbor 
tight-binding model, plus on-site disorder and contacts.  Thereby more subtle electronic structure and many-body effects were 
neglected. From a quantum chemical point of view the  $\pi$-electron  distribution and geometric aspects, such as bond length and 
hexagon area alternations at and near the edges, for sure should be more seriously taken into account for narrow (aromatic) armchair
GNRs,\cite{WSSLM10} maybe by an effective third nearest neighbor hopping.\cite{ZG09}  While such couplings will significantly 
influence the band gap---and  hence the transport properties of clean GNRs---Anderson localization, if present owing to bulk disorder, 
should be less affected. Coulomb interaction effects are particularly important for zigzag GNRs, where spin polarised edge states 
have been predicted.\cite{YPSCL07} An equal-footing treatment of disorder and Coulomb correlations in low-dimensional 
many-particle systems has turned out to be extremely difficult. Modeling possible magnetic properties/functionalities of (disordered) zigzag 
GNRs by an ad hoc spin-polarizing field might help to keep the problem tractable.\cite{GALMW07} 
These will be interesting directions for continuative work.

\section*{Acknowledgments}
This work was supported by the Deutsche Forschungsgemeinschaft through 
the priority programmes 1459 `Graphene' and 1648 `Software for Exascale Computing'.
Part of this work was performed at the Center for Integrated Nanotechnologies at Los Alamos National
Laboratory (DOE Contract DE-AC52-06NA25396).

%\bibliography{./ref} 

\bibliographystyle{apsrev}

\bibliography{ref}

\end{document}